\documentclass[11pt,a4paper]{article}
\usepackage{amsmath,amssymb}
\usepackage{epsfig,graphicx}

\begin{document}

\title{\bf CMB anisotropy induced by a moving straight cosmic string}
\author{O.~S.~Sazhina$^a$\footnote{{\bf e-mail}: tedeshka@mail.ru},
M.~V.~Sazhin$^{a}$\footnote{{\bf e-mail}: moimaitre@mail.ru},
V.~N.~Sementsov$^{a}$,
M.~Capaccioli$^{b,c}$,
G.~Longo$^{b}$, \\
G.~Riccio$^{b}$,
G.~D'Angelo$^{b}$
\\
$^a$ \small{\em Sternberg State Astronomical Institute of Moscow State University} \\
\small{\em 119992 Moscow, Universitetskiy pr., 13} \\
$^b$ \small{\em University of Napoli Federico II, Departament of Physical Sciences} \\
\small{\em 80126 Naples, via Cinthia, 6} \\
$^c$ \small{\em VSTceN-INAF} \\
\small{\em 80131 Naples, via Moiariello, 16}
}
\date{}
\maketitle

\begin{abstract}

We showed that the part of strings could be detected by optical method is only $20\%$ from the total available amount of such objects, therefore the gravitational lensing method has to be "`completed"' by CMB one. We found the general structure of the CMB anisotropy generated by a cosmic string for simple model of straight string moving with constant velocity. For strings with deficit angle 1-2 arcsec the amplitude of generated anisotropy has to be $15-30 \mu K$ (the corresponding string linear density is     $G \mu \propto 10^{-7}$ and energy is GUT one, $10^{15}$GeV). To use both radio and optical methods the deficit angle has to be from 0.1 arcsec to 5-6 arcsec. If cosmic string can be detected by optical method, the length of corresponding brightness spot of anisotropy has to be no less than 100 degrees.
\end{abstract}

\section{Introduction}

Cosmic strings are linear topological defects could form in the early Universe during face transitions. These objects have been introduced in theoretical cosmology by \cite{k}, \cite{z}, \cite{v}. 

Among all possible types of topological defects, cosmic strings are particularly interesting and their existence finds support in superstring theories, both in compactification models and in theories with extended additional dimensions \cite{h}, \cite{vs}, \cite{dk}. The energy scale for cosmic strings is the scale of GUT or less. The energy scale for fundamental strings is Planck scale. Such heavy strings do not exist in our Universe today, and cannot  have played any role in cosmological evolution except in the first few Planck times. Now we know models with large compact dimensions, in which  the string scale may be much lower, down to the GUT scale or even less. Branes, which now play key role in superstring theory, can collide and generate cosmic strings.

At the current moment there are no observational evidences to prove the existence of cosmic strings, neither to disprove it. The status of cosmic strings as theory related with observational data is based on the possibility to restrict  number, linear density and spatial distribution of such objects. These topological defects can influence on the CMB (Cosmic Microwave Background) anisotropy: according to WMAP 5-years data they are not primary source of primordial density perturbations and not responsible for large scale structure formation, therefore their density has to be hard restricted and be small enough.  From theoretical point of view all characteristics of cosmic strings are resulted mostly by extensive multiparameter computer simulations of string networks. But it is very important to note that these simulations are oriented to search string networks only, not individual strings as we now proposed in our work, because no theory tells us how many strings could exist in the Universe.

The modern methods of cosmic string detection can be divided in three main parts: 
\begin{itemize}
	\item by optical surveys looking for gravitational lensing events,
	\item by radio surveys investigating the structure of CMB anisotropy,
	\item looking for model depended and rare features, as gravitational radiation from string loops and from straight string, interactions of strings and black holes, decay of heavy particles emitted by string, interaction of two strings etc.
\end{itemize}
The superstring theory permits existence of cosmic strings in wide range of their parameters: with linear density varying from the scale of electroweak unification to GUT scales and with velocity from zero to speed of light. Also there are permissible strings possessed of curvature and loop like structures. All strings share two properties which are model independent: the extremely long cosmological length and a negligibly small cross-section. The investigation of the structure of CMB anisotropy looking for string footprints and search for gravitational lensing events induced by strings are so attractive because these effects should exist for all theoretically possible cosmic strings. 

Gravitational lensing events should have particularly and easy distinguishing with high resolution instruments (as HST) features. Cosmic string always generates gravitational lensed pairs of images of background sources, therefore direct detection of such pairs should be direct proof of cosmic string existence. On 11th January 2006 the russian-italian group (\cite{s0},\cite{s1}) received and completely analyzed the data from HST on the object CSL-1 which was unique candidate to be produced by a cosmic string, and it was proved that this is not a case of cosmic string lensing.  A base of this experience it was elaborated the technique of cosmic string search through gravitational lensing using ultra-deep optical surveys with high resolution instruments (see, for example, \cite{s1}). Analizing the results we found that the gravitational lensing method possesses very serious lack of sky covering both in terms of area and depth. The method of searching of gravitational lensing events is optical one. Modern optical surveys cover only $1/6$ part of the whole sky and their red shift depth is not more than z=7. Therefore, approximately only $3\%$ of possible strings could be detected by this method (if we suppose simple homogeneous distribution of straight strings in the Universe). Taking into account the predicted number of cosmic strings it was estimated that gravitational lensing events are very rare ones: the number of so-called "chains" of gravitational lensed pairs is from 0.3 up to 3 in the most optimistical models. This estimation can explain unsuccessful previous experience and shows that additional methods of cosmic string search are urgently required because even utilization of ultra deep galaxy surveys is not enough.

The investigation of CMB anisotropy seems to be the most fruitful and natural in cosmic string search, because of two following reasons. Firstly, this method operates with one of the model independent  properties of cosmic strings, their possibility to form conical space-time. Secondly, the red shift depth of radio surveys is z=1000 and covers the whole sky. Therefore using the WMAP 5-years data on CMB anisotropy we can make search of cosmic strings in all volume inside the surface of last scattering and could detect $100\%$ strings, not $3\%$ as in optical surveys, or well-groundedly disprove existence of such objects for wide range of their parameters. 

In the present work it was elaborated the method of searching for cosmic strings based on analysis of the CMB anisotropy. Moving straight cosmic string was shown to generate distinctive structures of enhanced and reduced temperature fluctuations on the surface of last scattering. It was analized the conditions under which a cosmic string could be detected by both CMB anisotropy and gravitational lensing. 

\section{Cosmic string as cosmological object}

Cosmic string is the topological defect which could form in early Universe in phase transitions. The Universe which contains a string is not flat. It is formed the conical space-time, where the cusp of the cone coincides with the position of the  string in current moment. The complete turn in a plane perpendicular to the string gives the total angle smaller than $2\pi$. This way, for any circle which goes around the cone cusp the ratio of circle length to the circle diameter is not equivalent to $\pi$. To put this idea more clear, let us introduce the system of lagrangian coordinates $O \xi \eta \zeta$. In cylindrical coordinates
\begin{eqnarray*}
\rho & = & \sqrt{\xi^2+\eta^2} \\
\xi  & = & \rho \cos \phi \\
\eta & = & \rho \sin \phi,
\end{eqnarray*}
\noindent where $O \zeta$-axis coincides with string, the first part of string metric is Minkowskian one and the second part is logarithmic function of the distance and depends on linear string density $\mu$: 
\begin{eqnarray}
\label{m} 
g_{\alpha \beta}   & = & \eta_{\alpha \beta} + h_{\alpha \beta},
\end{eqnarray}
\begin{eqnarray*} 
\eta_{\alpha \beta}& = & diag \{ 1,-1,-1,-1 \}, \\
 h_{\alpha \beta}  & = & 8 G \mu \ln \frac{\rho}{\rho_0} diag \{ 0,1,1,0 \}.
\end{eqnarray*}
If distance from the string goes to infinity, then the second term in string metric also diverges. 
If we introduce the following coordinates' transformation:
\begin{eqnarray*}
\hat{\tau} & = & \tau \\
\hat{\xi}  & = & \xi + 4 G \mu \biggl( 1-\ln \frac{\rho}{\rho_0} \biggr) \xi + 4 G \mu \phi \eta \\
\hat{\eta}  & = & \eta + 4 G \mu \biggl( 1-\ln \frac{\rho}{\rho_0} \biggr) \eta - 4 G \mu \phi \eta \\
\hat{\zeta} & = & \zeta
\end{eqnarray*}
the metric (\ref{m}) becomes Minkowskian with cut as function of deficit angle $D$, which depends on string linear density $\mu$:
\begin{eqnarray*}
\hat{g}_{\alpha \beta} = \eta_{\alpha \beta} = diag \{ 1,-1,-1,-1 \}, \\
0 \leq \phi \leq 2 \pi - D, \\
D = 8 \pi G \mu. 
\end{eqnarray*}

We used the cosmological model of Friedman Universe with $\Lambda$-term
\begin{eqnarray}
ds^2 & = & c^2dt^2 - a^2(t)(d\xi^2 + d\eta^2 + d\zeta^2),
\label{f0}
\end{eqnarray}
\begin{eqnarray}
\biggl( \frac{\dot{a}(t)}{a(t)} \biggr)^2 & = & \frac{8 \pi}{3} G \rho_c \biggl[ \Omega_M \biggl( \frac{a(t_0)}{a(t)} \biggr)^3 + \Omega_{\Lambda} \biggr],
\label{f1}
\end{eqnarray}
where $a(t_0)=1$, $\Omega_M=0.28$, $\Omega_{\Lambda}=0.72$. The Universe density is equal to the critical density 
$$
\rho_c = \frac{3}{8\pi G}H_0^2,
$$
with $H_0 = 73$km/s/Mpc (\cite{wmap1}, \cite{wmap2}).

\section{General structure and amplitude of anisotropy induced by a cosmic string}

As the Universe expands the temperature of the plasma falls down the recombination temperature and the formation of neutral hydrogen becomes energetically favorable. The photons began to propagate freely in space. This primordial radiation formed the surface of last scattering. There is a small anisotropy in its temperature, of order of $10^{-5}$.	The amplitude vibration of the surface of last scattering is a function of multipole number. In the low-multipole part of the angular spectrum the Standard Cosmological Model is not in very good agreement with observational data. It can be explained by non-trivial topology of the Universe, or by one or several cosmic strings. As we mentioned above, there are several methods to cosmic string detection, but not primary ones and model depended. We concentrate our efforts in CMB anisotropy method and will compare it with independent optical method of searching of gravitational lensing events.

We investigate a single straight infinite cosmic string moving with constant velocity in the homogeneous and isotropic background (\ref{f0}) - (\ref{f1}). A moving string could generate CMB anisotropy according a simple Doppler mechanism. Our aim is to estimate the amplitude of anisotropy by a single string in the homogeneous and isotropic background and to clarify its general structure and properties. For our purpose, from the point of view of cosmic string detection, the anisotropy due adiabatic perturbations being of the order of $10^{-5}$ is just a noise.  

The motion from the surface of last scattering to observer manifests itself as a frequency shift for one photon and as temperature fluctuation for an ensemble of photons with planckian spectrum: 
\begin{eqnarray*}
\frac{\delta T}{T} = \frac{E_o - E_e}{E_e} = \frac{k^{\alpha}u_{\alpha(o)} - k^{\alpha}u_{\alpha(e)}}{k^{\alpha}u_{\alpha(e)}}, \\
u^{\alpha} = \gamma \biggl \{ 1, \frac{v}{c} \cos(\psi - 4G \mu \phi_e), \frac{v}{c} \sin(\psi - 4G\mu \phi_e),0  \biggr\}, \\
\tan \phi_e = \frac{\sin \phi \sin \theta}{\cos \phi \sin \theta - \rho_{os}/\rho_{oe}},
\end{eqnarray*}
where $E_o$ is the photon self-energy in the observer reference frame, $E_e$ is the photon self-energy in the emitter (element on the surface of last scattering) reference frame, $k^{\alpha} = \omega/c(1,\bf{n})$ is wave-vector from observer to emitter, $u^{\alpha}$ is 4-velocity of a photon, $\psi$ is an angle between string velocity  and line connecting emitter and observer, $\phi$ and $\theta$ are angles in the spherical coordinate system with center in the observer, $\rho_{os}$ and $\rho_{oe}$ are distances from observer to string and from observer to emitter respectively (see for details \cite{ss}).

Following the standard procedure to calculation of the amplitude of string anisotropy (\cite{st}):
\begin{eqnarray*}
\frac{\delta T}{T} & \approx & 8\pi G \mu \gamma \frac{v}{c}, 
\end{eqnarray*}
where as usual $\gamma=1/\sqrt{1-v^2/c^2}$, and $v$ is projection of string velocity on the line perpendicular to line connecting emitter and observer. 

According our model, an appeared anisotropy induced by single cosmic string represents a sequence of zones of decreased and increased temperature: the cold spot in front of moving string, then the step-like jump and appearance of hot spot and than a cold spot follows again (see Fig.1). The amplitude of these discontinuities and their extension depend on position of the string with respect of an observer, on the string velocity and its direction, and on the string linear density. But the structure always remains the same. The simulations showed, that for cosmic string with deficit angle from 1 to 2 arcsec (these values correspond to linear density of GUT energy scale) the amplitude of generated anisotropy would be from 15 to 30 $\mu$K. The upper bound on string deficit angle is $D = 6''$, which corresponds to $\delta T/T \approx 100 \mu$K. The bound appears because in this case the induced string anisotropy is compared with anisotropy due adiabatic fluctuations and therefore has to be detected directly. The lower observational bound which is defined by angular resolution of optical telescopes like HST (Hubble Space Telescope) is $D = 0''.1$ , which corresponds to $\delta T/T \approx 1.5 \mu$K. So-called "`superlight"' strings could exist but can not be detected using the methods discussed. 

Also we have to take into account the effect of time retardation over the string extension (\cite{v86}). If an infinite straight string moves past the observer, he sees its distant parts with a retardation, so that the string appears to be curved. The result that the anisotropy spots should possesses the curvature too (\cite{s_pr}, in preparation).

The length of the spot (see Fig.1) depends on how long part of the string is inside the surface of last scattering. Therefore, the minimum of length of the spot, which can be in principle observed in optical surveys, at least 100 degrees, due the fact that optical depth is not more than z=7. Taking into account these estimations, we also can give an explanation why all attempt to find a cosmic strings through gravitational lensing effects were fallen. In average, only $20\%$ of all cosmic strings could be seen in optic ($20\%$ in the whole sky and only $3\%$ by current optical surveys which cover only $1/6$ part of the sky). This fact demonstrates that our alone string search seems to be more effective than the string network simulations.

\section{Conclusions and plans for future investigations}

Cosmic string could produce distinctive features in distribution of CMB: the temperature would have step-like discontinuities. It was analyzed in details the geometry of such step-like structures, depending on the distance to a cosmic string and its velocity. We analyzed the simplest model of string dynamic with isotropy CMB. 

An appeared anisotropy induced by single cosmic string represents a sequence of zones of decreased and increased temperature: the cold spot in front of moving string, then the step-like jump and appearance of hot spot and than a cold spot follows again (see Fig.1 --- Temperature distribution over the sky sphere (Molweide projection). The equator coincides with the horizontal line. The string lies along the axis connecting the poles. The angle $\phi$ is measured from the string front from right to left. There is a small cold spot in front of the string; a sharp temperature jump occurs at the front followed by an extended hot spot, which again gives way to an indistinct cold spot. The temperature distribution is typical of a moving string and is virtually independent of its parameters, only the spot width and the temperatures at a maximum and local minima change.). The amplitude of these discontinuities and their extension depend on position of the string with respect of an observer, on the string velocity and its direction, and on the string linear density. But the structure always remains the same. 

For a cosmic string with deficit angle from 1 to 2 arcsec the amplitude of generated anisotropy would be from 15 to 30 $\mu$K. It was obtained the structure and amplitude of expected anisotropy induced by cosmic string as function of its parameters. It was shown that in order to detect cosmic strings by both methods of gravitational lensing and CMB anisotropy the range of deficit angle has to be from 0.1 arcsec to 6 arcsec, which covers the huge amount of theoretically predicted strings. If cosmic string can be detected by optical method, the length of corresponding brightness spot of anisotropy has to be no less than 100 degrees.

Therefore our current investigations achieved the level which can certainly indicate the direct way to find cosmic strings in the whole Universe for wide range of string parameters or for the first time prove that such topological defects do not exist. According to superstring approach, cosmic strings are the most preferable objects to be formed in early Universe among other topological defects. There are also deep connections between cosmic strings and brane-world scenario based on superstrings. Long superstrings - macroscopic fundamental strings - may be stable and may appear at the same energy as GUT scale cosmic strings. They could have been produced in the early Universe and then grown to macroscopic size with the expansion of the Universe.

But more work is required. To really detect the string on the sky we have to distinguish the CMB anisotropy induced by scalar density perturbation from the cosmic string one; the problem is a low signal to noise ratio. The amplitude of noise (adiabatic perturbations) is larger then expected signal from cosmic strings. To detect a cosmic string with parameters set by GUT we have to detect anisotropy with signal to noise ratio much less then unity. For detection of the signal which is less then noise we intent to apply two method: wavelet and curvelet analysis and detection by optimal filtering.

\section*{Acknowledgments}
Sazhina O.S. and Sazhin M.V. acknowledge the Organizing Committee of QUARKS2008 to be kindly invited to present a talk. Sazhina O.S. acknowledges the RF President Grant MK-2503.2008.2. The work was supported in part by Grant RFFI-07-02-01034a.

\label{lastpage}

\end{document}